\documentclass[%
twocolumn,
superscriptaddress,
showpacs,
preprintnumbers,
 amsmath,
 amssymb,
 aps,
 pra,
floatfix,
]{revtex4-1}

\usepackage{graphicx}
\usepackage{adjustbox}
\usepackage{bbm}
\usepackage{bm}
\usepackage{csquotes}
\usepackage[colorlinks=true, citecolor=blue, linkcolor=blue, urlcolor=blue]{hyperref}
\usepackage[usenames,dvipsnames]{xcolor}

\usepackage{todonotes}

\renewcommand{\d}{\mathrm d}
\newcommand{\im}{\mathfrak i}
\newcommand{\omegad}{\omega_\d}

\newcommand{\tp}{t_\mathrm{p}}
\newcommand{\sigp}{\sigma_\mathrm{p}}

\newcommand{\defeq}{:=}
\renewcommand{\vec}{\boldsymbol}
\newcommand{\ket}[1]{|#1\rangle}

\newcommand{\C}{\mathrm 0}
\newcommand{\X}{\mathrm 1}

\newcommand{\aX}{\hat a^{\X}}

\newcommand{\aC}{\hat a^{\C}}
\newcommand{\aCd}{\hat a^{\C \dagger}}
\newcommand{\aXC}{\hat a^{\alpha}}

\begin{document}

\title{Resonant laser excitation and time-domain imaging of chiral topological polariton edge states}

\author{Damian Hofmann}
\affiliation{
    Max Planck Institute for the Structure and Dynamics of Matter,
    Luruper Chaussee 149, 22761 Hamburg, Germany
}

\author{Michael A. Sentef}
\affiliation{
    Max Planck Institute for the Structure and Dynamics of Matter,
    Luruper Chaussee 149, 22761 Hamburg, Germany
}
\affiliation{
    Institute for Theoretical Physics, University of Bremen,
    Otto-Hahn-Allee 1, 28359 Bremen, Germany
}

\date{\today}

\begin{abstract}
    We investigate the dynamics of chiral edge states in topological polariton systems under laser driving. Using a model system comprised of topolgically trivial excitons and photons with a chiral coupling proposed by Karzig \textit{et al.}~[Phys.~Rev.~X 5, 031001 (2015)], we investigate the real-time dynamics of a lattice version of this model driven by a laser pulse. By analyzing the time- and momentum-resolved spectral function, measured by time- and angle-resolved photoluminescence in analogy with time- and angle-resolved photoemission spectroscopy in electronic systems, we find that polaritonic states in a ribbon geometry are selectively excited via their resonance with the pump laser photon frequency. This selective excitation mechanism is independent of the necessity of strong laser pumping and polariton condensation. Our work highlights the potential of time-resolved spectroscopy as a complementary tool to real-space imaging for the investigation of topological edge state engineering in devices.
\end{abstract}

\maketitle
\section{Introduction} \label{sec:intro}

Light-matter interactions and nonequilibrium dynamics are unifying themes connecting solid state physics, quantum optics, and atomic and molecular physics. The prospect of controlling properties on demand in quantum materials by coupling them to laser fields has spurred enormous activity recently. Experimental progress in time-resolved spectroscopy has allowed researchers to obtain detailed information about the ultrafast dynamics of laser-excited quantum materials  \cite{giannetti_ultrafast_2016,basov_towards_2017,cavalleri_photo-induced_2018}. This is particularly relevant for the investigation of light-induced changes in material properties on ultrafast time scales. Notable examples are Floquet topological states of matter \cite{oka_photovoltaic_2009,lindner_floquet_2011,kitagawa_transport_2011,wang_observation_2013,mahmood_selective_2016,usaj_irradiated_2014,dehghani_dissipative_2014,sentef_theory_2015,claassen_all-optical_2016,hubener_creating_2017,mciver_light-induced_2020}, also demonstrated for cold atoms in optical lattices \cite{jotzu_experimental_2014,aidelsburger_measuring_2015}, ultrafast modifications of effective interactions and their consequences for the emergent material properties \cite{singla_thz-frequency_2015,dutreix_dynamical_2017,kennes_transient_2017,sentef_light-enhanced_2017,tancogne-dejean_ultrafast_2018,topp_all-optical_2018,golez_dynamics_2019,buzzi_photo-molecular_2020}, and cavity material engineering with quantum fluctuations of light \cite{laussy_exciton-polariton_2010,cotlet_superconductivity_2016,kavokin_excitonpolariton_2016,hagenmuller_cavity-enhanced_2017,sentef_cavity_2018,rosner_plasmonic_2018,schlawin_cavity-mediated_2019,mazza_superradiant_2019,curtis_cavity_2019,kiffner_manipulating_2019,hagenmuller_enhancement_2019,allocca_cavity_2019,rokaj_quantum_2019,latini_cavity_2019,forg_cavity-control_2019,thomas_exploring_2019}.
The theoretical description of these examples of light-induced states of matter usually involves electronic single-particle excitations, which are fermionic in nature. By contrast, polaritonic systems can often be described by a purely bosonic theory, for example in polaritonic condensates \cite{byrnes_excitonpolariton_2014}. In particular, topological edge states \cite{hasan_colloquium_2010,qi_topological_2011} in those systems are often well described already by an effective semiclassical approach, which simplifies theoretical calculations.
This has led to a variety of proposals for realizing topological polaritons, for example in photonic \cite{haldane_possible_2008,wang_observation_2009,rechtsman_photonic_2013,hafezi_imaging_2013}, acoustic \cite{yang_topological_2015,peano_topological_2015,fleury_floquet_2016}, and mechanical systems \cite{nash_topological_2015,susstrunk_observation_2015}. Other work providing evidence for the growing range of polaritonic platforms potentially hosting useful edge states includes studies of molecule-by-molecule assemblies and optically trapped ultracold atoms \cite{polini_artificial_2013} as well as semiconductor microcavities \cite{jacqmin_direct_2014,milicevic_edge_2015,sala_spin-orbit_2015}.

Recently, polaritonic systems hosting chiral topological edge states were proposed theoretically \cite{karzig_topological_2015,nalitov_polariton_2015,yuen-zhou_plexciton_2016} and measured experimentally \cite{klembt_exciton-polariton_2018}. For any practical application of chiral topological edge modes it is of key importance to be able to selectively populate these modes, and to track whether such selective population has been successful.
However, for example in the work by Karzig et al.~\cite{karzig_topological_2015}, the degree of control over the population of such edge mode in terms of tuning of relevant parameters was not discussed in much detail. In particular it was only noted in passing that the controlled population of the chiral edge mode by laser driving is enabled either by tuning the laser frequency into resonance with the topological polariton band gap or by focussing the laser spot to the edge in real space, or a combination thereof.

In the experimental work by Klembt et al.~\cite{klembt_exciton-polariton_2018} an exciton-polariton chiral topological system was realized in a honeycomb lattice with a magnetic field breaking time-reversal symmetry. It was shown by photoluminescence measurements that chiral edge modes could be populated by laser pumping. Due to the fact that this selective edge-mode population was only observed above a certain threshold power, it was argued that the formation of a polariton condensate was a prerequisite of such edge mode population. This condensate formation, in turn, requires a nonlinearity in the underlying Gross-Pitaevskii equation, stemming from a polariton-polariton interaction in the microscopic model.

Here we show by calculations for a topological polariton lattice model how real-space imaging and time-resolved spectroscopy go hand-in-hand in demonstrating selective edge mode excitations by laser driving. We demonstrate that selective excitation of chiral edge modes is possible without the necessity of polariton condensation due to nonlinearities, that is, without polariton-polariton interactions. However, in the absence of nonlinearities it is necessary that the pump laser frequency is tuned in resonance with the edge mode, that is, with the bulk topological band gap. Complementary real-space imaging then reveals the edge localization and chiral character of the populated modes.

This paper is organized as follows: In Section~\ref{sec:model}, the model is presented and its physics introduced. In Section~\ref{sec:dynamics}, we discuss how the dynamics of the laser-driven system is simulated and tracked through time-resolved spectroscopy. Section~\ref{sec:results} contains the results of these simulations, and Section~\ref{sec:discussion} contains a discussion and conclusions.

\begin{figure}[tbp]
    \centering
    \includegraphics[width=0.46\textwidth]{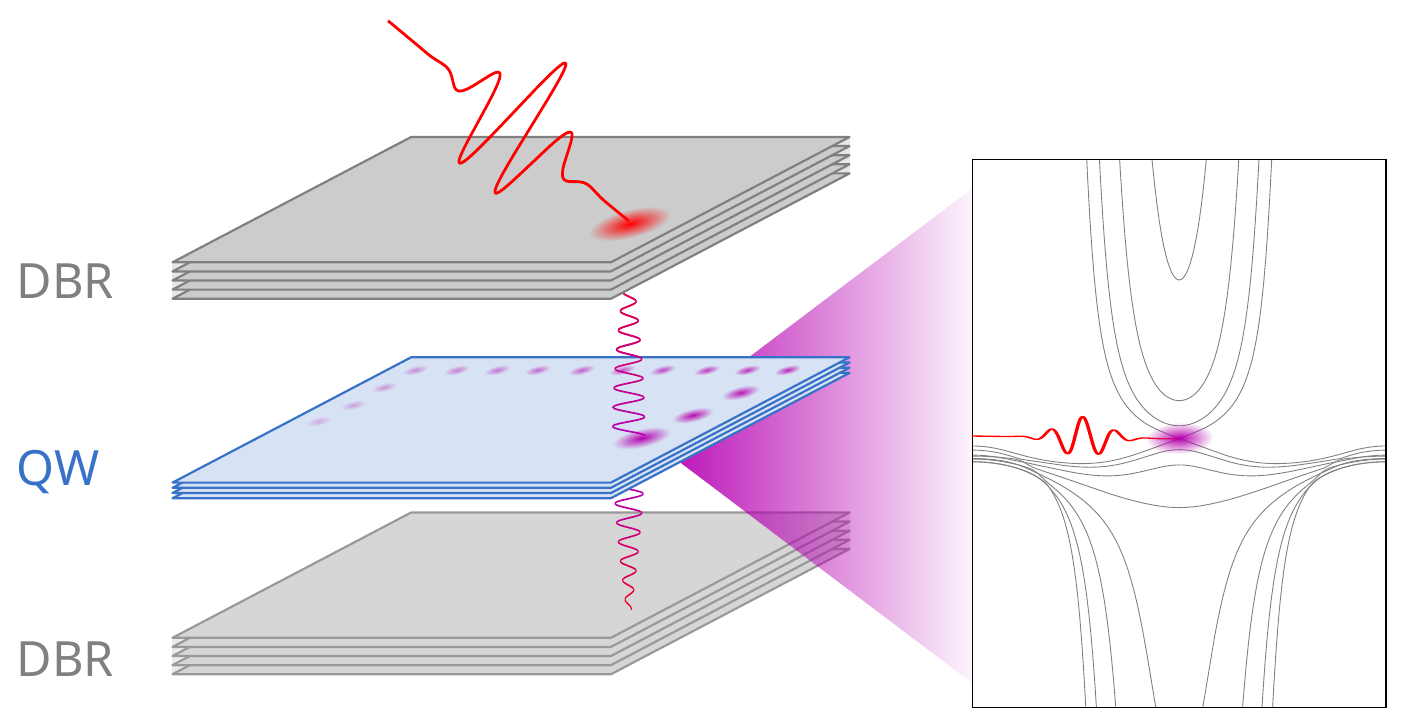}
    \caption{Illustration of a semiconductor cavity setup. An isolated quantum well (QW) is placed between two optical dynamical Bragg reflectors (DBR). The cavity photon modes are populated by optical pumping and couple to excitonic states within the QW plane via dipole interaction.
        The specific coupling under consideration here yields topological edge states which can be selectively populated.
    }
    \label{fig:1}
\end{figure}

\section{Model} \label{sec:model}

\begin{figure}[tbp]
    \centering
    \includegraphics[width=\columnwidth]{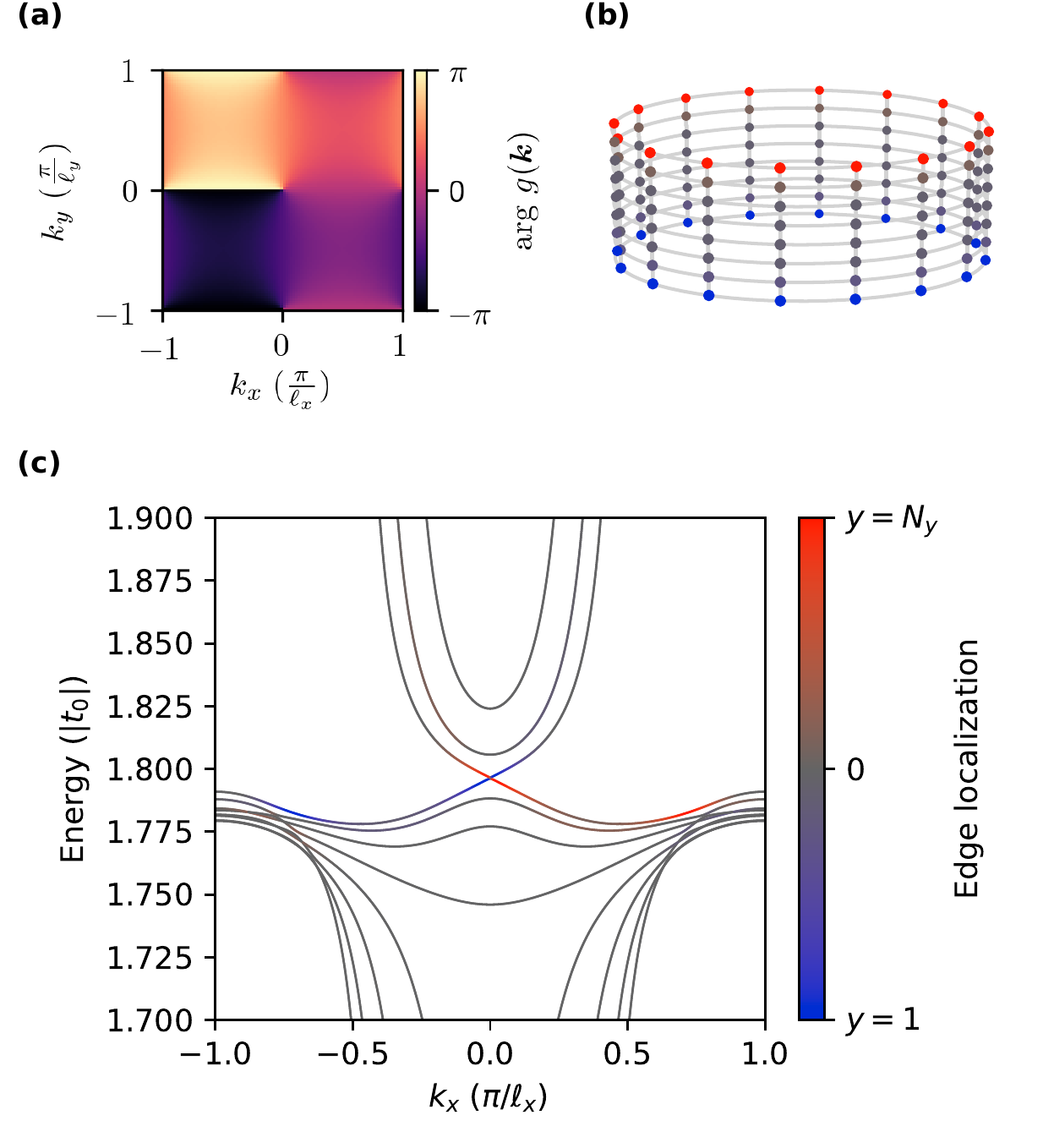}
    \caption{
        (a) Complex phase of the chiral exciton-polariton coupling [Eq.~\eqref{eq:coupling}]
        with a branch cut along the $\mathrm\Gamma = (0, 0)$ to $\mathrm X = (-\frac{\pi}{\ell_x}, 0)$ line.
        This is the phase-winding structure that leads to a non-trivial topological band structure.
        (b) Real-space lattice structure of a ribbon with periodic boundary conditions in $x$ and open boundary conditions in $y$ direction.
        (c) Tight-binding topological polariton band structure of the the Hamiltonian $\hat H_0$ [Eq.~\eqref{eq:H0}] on the ribbon geometry.
        The bands are colored according to their localization at the top or bottom boundary as indicated in panel (a).
        The model parameters are $t_\X = 0.002 |t_\C|,$ $g_0 = 0.2 |t_\C|,$ $\varepsilon_\C = -4 |t_\C|$, and $\varepsilon_\X = -1.792 |t_\C|$, with an effective photon hopping $t_\C < 0.$
    }
    \label{fig:2}
\end{figure}

Our setup is shown in Fig.~\ref{fig:1}. A quantum well (QW) harbors exciton modes that couple via dipole-dipole interactions to photon modes in a cavity, leading to the formation of exciton polaritons. Alternatively, the QW could just as well be replaced by monolayer transition metal dichalcogenides \cite{karzig_topological_2015,xiao_coupled_2012}. In the following we will assume that there is a chiral coupling between excitons and polaritons. This can be achieved by employing an external magnetic field to energetically select chiral electronic states over their time-reversed counterparts with opposite chirality, which has been suggested by Karzig et al. \cite{karzig_topological_2015} and experimentally realized by Klembt et al. \cite{klembt_exciton-polariton_2018}.
From these chiral electrons one obtains excitons with nonzero total angular momentum. These excitons then couple in a chiral fashion to the photon branch with opposite angular momentum compared to the exciton, whereas it couples in a nonchiral fashion to the photon branch with equal angular momentum.
Thus, as noted by Karzig et al., the chiral coupling can be viewed as a simple consequence of angular-momentum conservation.

In this work, we will focus our attention on a tight-binding model on a two-dimensional square lattice,
which is designed to feature non-trivial topology arising from the chiral coupling of two bosonic modes in analogy to this mechanism.
A simple coupling satisfying this requirement while at the same time being consistent with the lattice periodicity is given by
\begin{align}
    \label{eq:coupling}
    g(\vec k) & = g_0\left[ \sin(\ell_x k_{x}) + \im \sin(\ell_y k_{y}) \right]
\end{align}
with coupling strength $g_0 > 0$ and where $\ell_{x/y}$ denotes the respective lattice constant.
For one revolution of $\vec k$ around the origin, the coupling accumulates a phase of $2\pi$ [Fig.~\ref{fig:2}(a)].
This winding is the key ingredient leading to a non-trivial topology of the system.

In the presence of periodic boundary conditions in both directions, the full static Hamiltonian in momentum space representation has the form
\begin{align}
    \label{eq:H0-k}
    \hat H_0 = \sum_{\vec k \in \mathfrak L'} \begin{pmatrix}
        \hat a^\C_{\vec k} \\ \hat a^\X_{\vec k}
    \end{pmatrix}^\dagger
    \begin{pmatrix}
        \tau_\C(\vec k) & g(\vec k) \\ g^*(\vec k) & \tau_\X(\vec k)
    \end{pmatrix}
    \begin{pmatrix}
        \hat a^\C_{\vec k} \\ \hat a^\X_{\vec k}
    \end{pmatrix}
\end{align}
where the bosonic operators $\hat a^\C_{\vec k}$ ($\hat a^\X_{\vec k}$) annihilate a photon (exciton) with momentum $\vec k.$
Further, $\mathfrak L'$ denotes the discretized first Brillouin zone of the lattice and
\(
\tau_\alpha(\vec k) = 2 t_\alpha \left[ \cos(\ell_x k_{x}) + \cos(\ell_y k_{y}) \right] - \varepsilon_\alpha
\)
(for $\alpha \in \{0,1\}$) is the tight-binding dispersion with exciton/photon hopping $t_\alpha$ and $\varepsilon_\alpha$ an energy offset.
The static Hamiltonian has a band structure that corresponds to a tight-binding version of the continuum model presented by Karzig et al.~\cite{karzig_topological_2015}.
The momentum-dependence of the coupling $g(\vec k)$ leads to a non-zero Chern number $C_{\pm} = \mp 1$ of the upper (${+}$) and lower (${-}$) polariton band.
A more detailed explanation is given in the Appendix.

In order to study the real-space behavior of edge mode excitations, we introduce boundaries to the system by imposing open boundary conditions in $y$ direction, while keeping periodic boundary conditions in $x$ direction (so that the momentum $k_x$ is still a good quantum number).
This leads to  a ribbon geometry, as shown in Fig.~\ref{fig:2}(b).
Let $\mathfrak L$ denote this real-space lattice of size $N = N_x \times N_y$ and $\aC_i$ ($\aX_i$) the photonic (excitonic) field operator at site $i \in \mathfrak L$.
The real-space model corresponding to Eq.~\eqref{eq:H0-k} is given by
\begin{align}
    \label{eq:H0}
    \hat H_0 = \sum_{i \in \mathfrak L} \sum_{d\in\{0, d_x, d_y\}} \sum_{\alpha, \beta \in \{\C,\X\}}
    \!\!\! t_{\alpha\beta}(d) \hat a^{\alpha\dagger}_i \hat a^{\beta}_{i+d}
\end{align}
where $i + d_\nu$ is the index of the nearest neighbor of site $i$ in positive $\nu \in \{x,y\}$ direction.
The hopping is determined by the matrices
$
    \mathbf t(0) = \mathrm{diag}(-\varepsilon_\C, -\varepsilon_\X)
$
and
\begin{align}
    \mathbf t(d_x) & = \begin{pmatrix}
        t_\C & \im g_0 / 2 \\ \im g_0 / 2 & t_\X
    \end{pmatrix}\!,
                   &
    \mathbf t(d_y) & = \begin{pmatrix}
        t_\C & g_0 / 2 \\ -g_0 / 2 & t_\X
    \end{pmatrix}\!.
\end{align}
Note that $\mathbf t(-d_\nu) = \mathbf t^\dagger(d_\nu)$ so that $\hat H_0$ is Hermitian.
As above, the parameters $t_\alpha$ are the photon and exciton hopping amplitudes,
$\varepsilon_\alpha$ constant energy offsets, and $g_0$ the exciton-photon coupling strength.
For a specific choice of parameters, this model is equivalent to the Qi-Wu-Zhang \cite{qi_topological_2006} or half Bernevig-Hughes-Zhang model \cite{bernevig_quantum_2006},
which is well known as a simplified tight-binding model for the description of topological insulators \cite{asboth_short_2016}.
Bosonic transport in a variation of this model including a non-linear interaction term has been studied by Wei\ss{} \cite{weis_nonlinear_2017}.

Figure~\ref{fig:2}(c) shows the band structure of the Hamiltonian \eqref{eq:H0} obtained by a partial Fourier transform in $x$ direction.
As expected, the model possesses a topological gap which is only crossed by a pair of bands, the eigenstates of which are highly localized at the opposite boundaries of the ribbon.
We note that in this model, exciton and photon hopping need to be of opposite sign in order for the edge modes to be present, which can be achieved through a negative exciton effective mass in a real material.

\section{Dynamics}
\label{sec:dynamics}

The time-dependent driving is implemented by the operator
\begin{align}
    \label{eq:F}
    \hat F(t) & = f_0 \, \mathrm{e}^{-\im \omegad t} \, \aCd_{i_0} \; + \; \mathrm{H.c.}.
\end{align}
which, for simplicity, is taken to act directly only on the photonic mode $\alpha = 0$ at a single site $i_0$ located on the open side of the lattice boundary.
The parameter $f_0$ is the product of a dipole matrix element of the material under consideration and the electric field strength of the drive laser. The form of the driving is chosen to mimick the laser-driven dynamics induced by a laser that is focussed to the edge of the system and switched on at the initial time.
The full time-dependent Hamiltonian then has the form
\begin{align}
    \label{eq:hamiltonian}
    \hat H(t) = \hat H_0 + \hat F(t).
\end{align}

Our simulations are performed in the semi-classical limit, where the bosonic operators are replaced by scalar complex fields instead of using a full quantum-mechanical treatment.
This corresponds to restricting the possible quantum states to the coherent states $\ket{\psi}$ parametrized by the scalar complex field $\psi \in \mathbb C^{2N}$ satisfying $\aXC_i \ket{\psi} = \psi_i^\alpha \ket{\psi}.$
The time-dependent Hamiltonian~\eqref{eq:hamiltonian}, including the driving term $\hat F(t)$, is of second order in the creation and annihilation operators due to the absence of a polariton-polariton interaction term.
In this case, if the simulation is started in a coherent initial state, the semi-classical approach captures the exact quantum dynamics of the system.
The equations of motion for the fields $\psi_i^\alpha$ are ($\hbar = 1$)
\begin{align}
    \begin{split}
        \mathfrak i \, \frac{\d\psi_i^\alpha}{\d t} \; &= \!\!\!\!\! \sum_{\substack{\beta\in\{\C,\X\} \\ d\in\{0, \pm d_x,\pm d_y\}}} \!\!\!\!\! t_{\alpha\beta}(d) \psi_{i+d}^\beta
        \, + \, \delta_{i,i_0}\delta_{\alpha,0} \, f_0 \mathrm{e}^{-\im \omegad t}.
    \end{split}
\end{align}
The initial state is chosen to be the coherent polaritonic vacuum state with $\psi_i^\alpha(0) = 0$ for all $i$ and $\alpha$.

\begin{figure*}[tbp]
    \centering
    \includegraphics[width=0.9\textwidth]{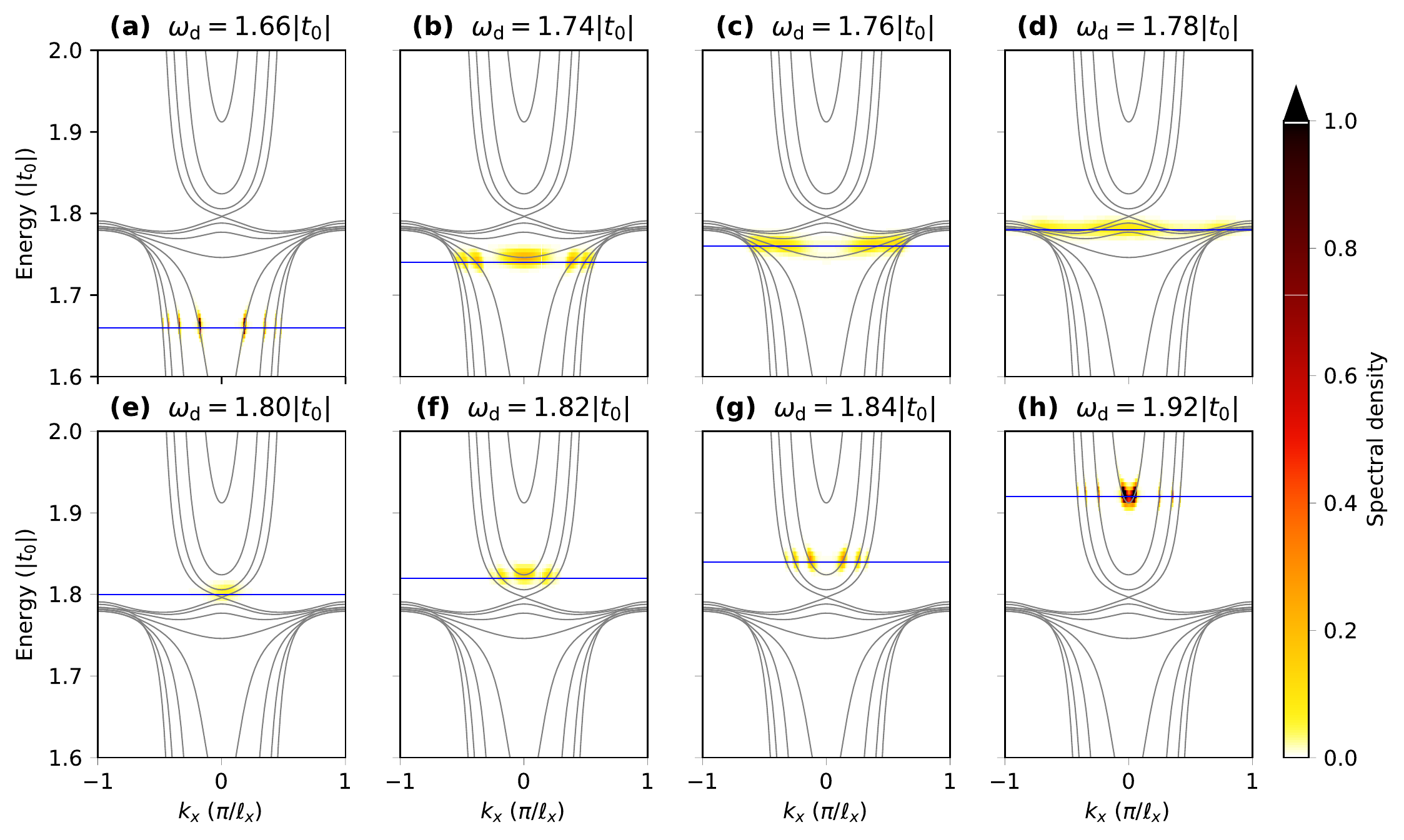}
    \caption{(a)--(h) Time- and momentum-resolved spectral density [Eq.~\eqref{eq:spec}] of the laser-irradiated system on a ribbon-shaped $N_x \times N_y = 256 \times 8$ lattice.
    The grey lines show the equilibrium band structure of the model \eqref{eq:H0}. The driving frequency $\omegad$ varies between subplots and is indicated by the blue line. The spectral density shows the excitation of the states resonant with the driving laser frequency.
    Here, the driving amplitude is $f_0 = 0.2|t_\C|$ and the spectral density has been computed at a probe time of $\tp = 400 |t_\C|^{-1}$ with $\sigp = 125 |t_\C|^{-1}.$
    The model parameters are the same as in Fig.~\ref{fig:2}.
    }
    \label{fig:3}
\end{figure*}

In the following we use double-time Green's functions in order to compute time-, momentum- and spatially resolved spectroscopy. 
This approach is generally applicable to nonequilibrium situations and capable of fully describing both transient and steady-state dynamics in driven systems.
For the specific case of a time-periodic driving field, we briefly note that one can, in principle, also obtain spectral information from a Floquet representation \cite{eckardt_high-frequency_2015}.
However, the Floquet framework has several limitations compared to our more general approach:
(i)~it only applies to a strictly time-periodic steady state, whereas we are specifically interested in the transient \enquote{switch-on} behavior and chiral propagation of topological edge modes;
(ii)~it is not straightforward to compute the population of Floquet states, in particular when there are no heat or particle baths attached to the system;
(iii)~it only captures the stroboscopic part of the time evolution with the period of the drive, but not the more complicated subcycle dynamics or micromotion.
In contrast, these limitations do not apply to double-time Green's function methods.

In the semi-classical limit, the double-time lesser Green's functions can be obtained directly from the bosonic fields as
\begin{align}
    \label{eq:Gless}
    \begin{split}
        G^<_{i\alpha;j\beta}(t_1, t_2)
        &\defeq -\im\langle \hat a^{\beta\dagger}_{j}(t_2) \hat a^{\alpha}_i(t_1) \rangle \\
        &= -\im\psi_i^\alpha(t_1) [\psi_j^\beta(t_2)]^*.
    \end{split}
\end{align}
This Green's function contains information about the propagation of a boson added to the system at time $t_2$ and removed from the system at time $t_1$. This can be used to extract both the spectrum of excitations as well as the occupation of such single-particle excited states.

From the Green's functions, we compute the time-resolved spectral density \cite{freericks_theoretical_2009}
\begin{widetext}
    \begin{align}
        \label{eq:spec}
        I(k, \omega, \tp) & = \operatorname{Im} \iint \d t_1 \d t_2 \, S_{\tp,\sigp}(t_1) S_{\tp,\sigp}(t_2) {\mathrm e}^{\im\omega(t_1-t_2))} \tilde G^<(k; t_1, t_2)
    \end{align}
\end{widetext}
with Gaussian probe shape
\begin{align}
    S_{\tp,\sigp}(t) & = \frac{1}{\sigp\sqrt{2\pi}} \exp\left[-\frac{(t - \tp)^2}{2\sigp^2}\right]
\end{align}
where $\tp$ is the probe time and $\sigp$ the Gaussian width given by the temporal duration of the probe laser pulse \cite{freericks_theoretical_2009}. The time resolution is thus determined by the shape function. As was shown in the context of electronic Green's functions \cite{PhysRevX.3.041033}, this time resolution comes at the expense of spectral resolution, and vice versa, due to Heisenberg's uncertainty principle. For our purposes, we are mainly interested in spectral information.
Therefore, the temporal duration of the probe pulse is chosen to be sufficiently large to be able to resolve the relevant spectral features.
The reduced lesser Green's function used in Eq.~\eqref{eq:spec} is
\begin{align}
    \tilde G^<(k_x; t_1, t_2) & = \mathrm{Tr}_{\alpha,y} \mathcal F_{x \to k_x} [G_{i_{x,y},\alpha;i_{x,y},\alpha}^{<}(t_1,t_2)],
\end{align}
which is Fourier-transformed along the periodic ribbon direction $x$ (denoted by $\mathcal F_{x \to k}$), in order to reveal momentum-resolved information along this direction, and traced over $y$ direction. We also trace here over the photon and exciton index, i.e., we compute the exciton-polariton spectral density. In principle, it is easily possible to resolve the spectral contributions of excitons and photons separately, but this will not be crucial for our analysis of edge localization of pumped modes below.

\section{Results}
\label{sec:results}

In Fig.~\ref{fig:3} we present the time- and momentum-resolved spectral functions, mimicking time-domain photoluminescence spectra, of a continuously irradiated ribbon of size $N_x \times N_y = 256 \times 8$ with laser focused to a lattice site $i_0$ at the lower edge ($y=0$) of the ribbon. The driving frequency of the external laser is varied to be below [Fig.~\ref{fig:3}(a-d)] the topological band gap, within the gap region [Fig.~\ref{fig:3}(e)], and above the gap [Fig.~\ref{fig:3}(g-h)]. As can be seen from the time-resolved spectral density, the polariton branches are selectively occupied by the resonant laser excitation, which shows that the external driving frequency is the main tuning knob for populating polariton branches in the absence of polariton-polariton interactions and associated mechanisms for polariton condensation.

\begin{figure}[tbp]
    \centering
    \includegraphics[width=\columnwidth]{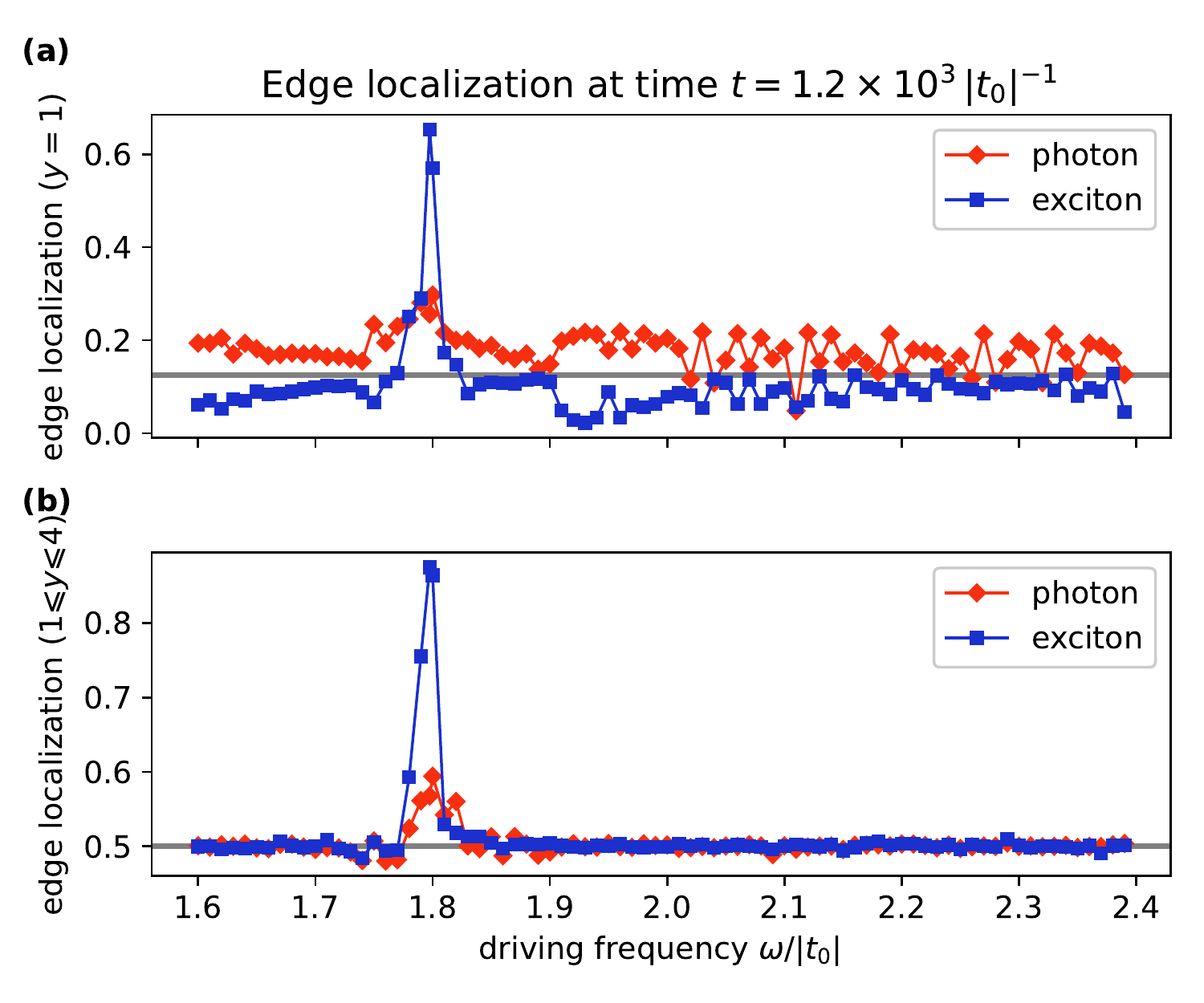}
    \includegraphics[width=\columnwidth]{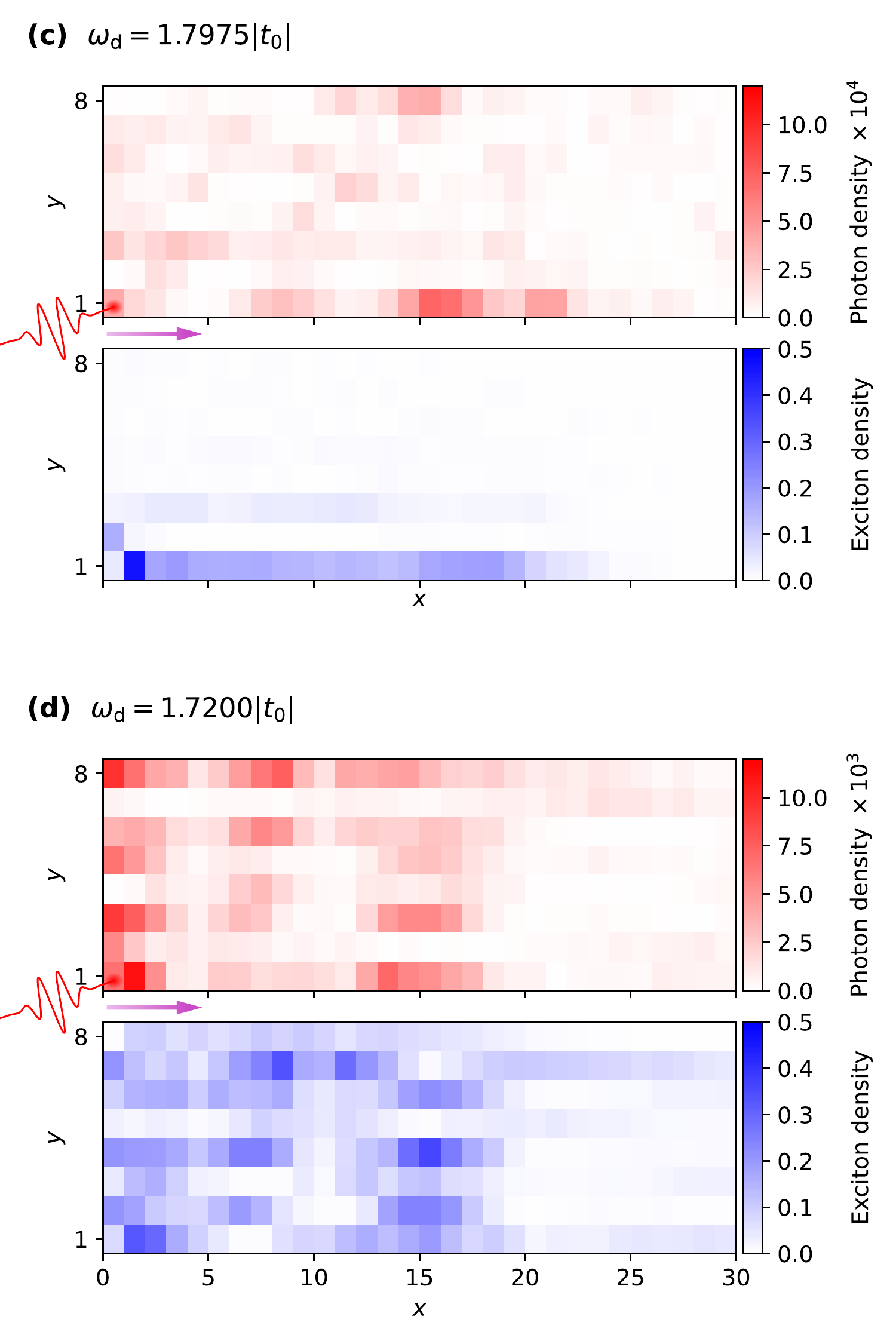}
    \caption{
    (a)--(b) Fraction of the total population located at the lowest layer $y=1$ (top panel) and within the lower half of the system $1 \leqslant y \leqslant 4$ (bottom panel). The horizontal grey lines indicate the fraction of the population located in the same region for a uniform distribution over the full $256 \times 8$ lattice.
    (c)--(d) Real-space density of the photon and exciton field at time $t = 1.2 \times 10^3 |t_0|^{-1}$ for resonant (panel c) and off-resonant (panel d) driving frequency.
    The driving field is localized at the bottom right corner of the displayed region.
    The direction of chiral propagation of the polariton modes is indicated by the purple arrow.
    }
    \label{fig:4}
\end{figure}

Finally, we analyze in Fig.~\ref{fig:4} the edge localization of the light-induced states through real-space analysis of the light-induced populations.
Fig.~\ref{fig:4}(a) shows the localization at the lower edge, i.e., the ratio of intensity at $y=1$ integrated (along the $x$ direction) and the total intensity in the system at a given point in time. A clearly resonant behavior is observed when the driving frequency matches the edge state energy of $\approx 1.8 |t_C|$. The degree of localization is even more pronounced for the excitonic component of the polariton wavefunction. In Fig.~\ref{fig:4}(b) we analyze the localization integrated over the lower half $1 \leq y \leq 4$ of the ribbon, which consistently shows same the resonant behavior. In Fig.~\ref{fig:4}(c) we show representative real-space images for the photon and exciton densities in the on-resonance case. Here, besides the already discussed localization at the lower edge, one can also observe the chiral propagation of the edge mode from left to right as indicated by the arrow. By clear contrast, in Fig.~\ref{fig:4}(d) both the edge selectivity and chirality of the laser-pumped polariton populations are absent. These real-space images thus provide complementary information to the spectroscopic pictures presented in Fig.~\ref{fig:3}.

\section{Discussion}
\label{sec:discussion}

In summary we have shown lattice model simulations for the temporally and spatially resolved dynamics of chiral topological exciton polaritons. We have demonstrated the selective excitation of chiral edge modes provided that the external driving frequency is sufficiently closely tuned to the topological band gap.
Importantly, the focusing of the laser to the edge of the system is by itself not sufficient to selectively populate the edge mode. This is likely due to the fact that also bulk states spatially extend into the edge region, and these bulk states can be populated even by a laser that is focused to the edge. An interesting open question pertains to the role of polariton-polariton interactions and the question of the importance of condensation versus resonant excitation for the experimentally observed chiral edge modes \cite{klembt_exciton-polariton_2018}.

As an upshot from our calculations, it is straightforward to extend the formalism employed here to compute time-resolved spectroscopy in periodically driven many-body systems, both with continuous and ultrashort laser pulses. Specifically for continuous driving this is the realm of Floquet engineering \cite{bukov_universal_2015,holthaus_floquet_2015,eckardt_high-frequency_2015,oka_floquet_2019} of effective band structures, which is by itself a rapidly growing research field.
Moreover, the recent discussion of breakdown of conventional bulk-boundary correspondence in non-Hermitian systems due to gain, loss, and violation of reciprocity, also points to the importance of complementary spectroscopic and imaging techniques to reveal topological edge modes and their selective population \cite{yao_edge_2018,xiong_why_2018,helbig_generalized_2020}. It is also interesting to contrast the present discussion of complementary techniques for detecting topological edge states with electronic edge state spectroscopy and imaging in quantum materials, for example in topological insulators by means of time- and angle-resolved photoemission spectroscopy \cite{soifer_band-resolved_2019}. The interface between the different fields of quantum simulators, nonequilibrium quantum materials science, and polaritonic condensates promises many interesting applications for future quantum technologies.

\begin{figure*}[htbp]
    \includegraphics[width=\textwidth]{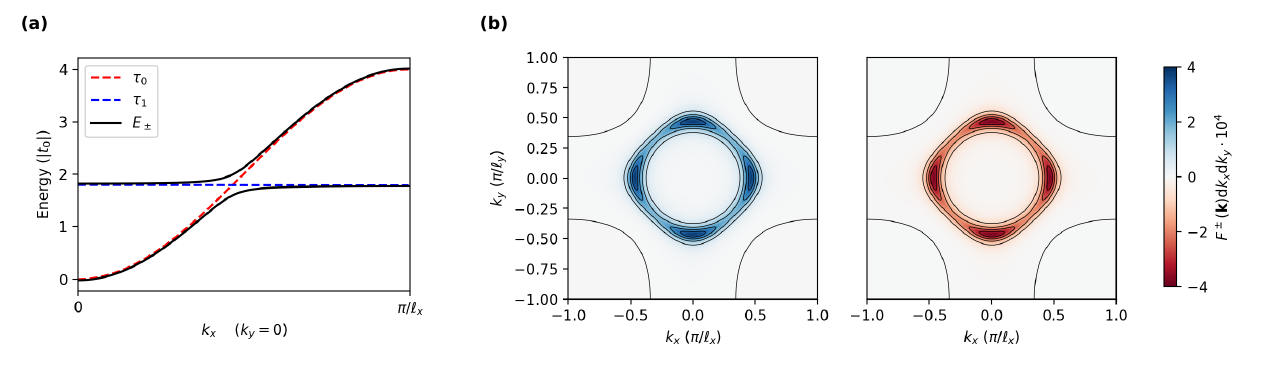}
    \caption{
    (a) Band structure of the Bloch Hamiltonian \eqref{eq:bloch} along a path from $\Gamma = (0,0)$ to $\mathrm X = (\pi/\ell_x, 0)$ in reciprocal space. Here, $E_\pm$ denotes the polariton band energies, while $\tau_{0}$ ($\tau_1$) is the uncoupled photon (exciton) dispersion. \\
    {(b)}~Discretized Berry curvature of the lattice model for the lower ($F^-$) and upper ($F^+$) band (left and right panel, respectively), computed using the method of Fukui et al. \cite{fukui_chern_2005} over the first Brillouin zone of the model \eqref{eq:bloch} discretized on a $512 \times 512$ grid.
    The model parameters are the same as in Fig.~\ref{fig:2}(c), except for the boundary conditions.
    The Chern number of the bands are $C_\pm = \sum_k F^\pm(k) \d k_x \d k_y = \mp 1$ for this case.
    }
    \label{fig:berry}
\end{figure*}

\begin{acknowledgements}
    We acknowledge helpful discussions with G.~Refael. We are particularly indebted to C.~Bardyn for sharing his Gross-Pitaevskii simulation code.
    Financial support by the DFG through the Emmy Noether program (SE 2558/2-1) is gratefully acknowledged.
\end{acknowledgements}

\appendix*

\section{Topological invariants of the tight-binding model}

\newcommand{\vcu}{\vec{\check u}}

The polariton lattice model described in Section \ref{sec:model} features a topological band structure characterized by non-zero Chern numbers, which we demonstrate here.
In the presence of periodic boundary conditions in both $x$ and $y$ direction, the lattice Hamiltonian \eqref{eq:H0} can be written in the momentum-space form of Eq.~\eqref{eq:H0-k}.
This Hamiltonian is of the general form
\begin{align}
    \hat H_0 = \sum_{\vec k \in \mathfrak L'} \begin{pmatrix}
        \hat a^\C_{\vec k} \\ \hat a^\X_{\vec k}
    \end{pmatrix}^\dagger
    \mathbf H(\vec k)
    \begin{pmatrix}
        \hat a^\C_{\vec k} \\ \hat a^\X_{\vec k}
    \end{pmatrix}
\end{align}
with the Bloch Hamiltonian $\mathbf H(\vec k)$ represented by a Hermitian $2 \times 2$ matrix.
The Pauli matrices together with the identity matrix $\boldsymbol{\mathbbm{1}}_2$ span the space of $2 \times 2$ Hermitian matrices and therefore the Bloch Hamiltonian can be expanded in this basis, giving
\begin{align}
    \label{eq:bloch}
    \mathbf H(\vec k) = \tau_+(\vec k) \boldsymbol{\mathbbm{1}}_2 + \vec d(\vec k) \cdot \boldsymbol{\hat \sigma}
\end{align}
where $\tau_+(\vec k) = \frac{1}{2}[\tau_0(\vec k) + \tau_1(\vec k)]$ is an energy offset and $\boldsymbol{\hat \sigma} = (\hat\sigma^x, \hat\sigma^y, \hat\sigma^z)^\top$ the vector of Pauli matrices.
The offset $\tau_+$ does not contribute to the Berry curvature and thus Chern number of the bands, which is determined solely by the gap vector $\vec d(\vec k).$
For the model studied here, the gap vector is given by
\begin{align}
    \vec d(\vec k) & = \begin{pmatrix}
        g_0 \sin(\ell_x k_x) \\
        -g_0\sin(\ell_y k_y) \\
        \tau_-(\vec k)
    \end{pmatrix}
\end{align}
where $\tau_-(\vec k) = \frac{1}{2}[\tau_0(\vec k) - \tau_1(\vec k)].$

The Chern number of the upper ($+$) and lower ($-$) polariton band are determined by the number of times the unit vector $\vcu(\vec k) = \vec{d}(\vec k) / \Vert \vec{d}(\vec k) \Vert$ wraps around the unit sphere when $\vec k$ is varied over the first Brillouin zone \cite{hasan_colloquium_2010,asboth_short_2016}.
Explicitly, it is given by
\begin{align}
    \label{eq:appx-chern}
    C_\pm = \pm \frac{1}{4\pi} \int \! \d^2 \vec k \, \left(\partial_{k_x} \vcu(\vec k) \times \partial_{k_y} \vcu(\vec k) \right) \cdot \vcu(\vec k).
\end{align}
For fixed $d_z(\vec k)$, $\vcu$ performs one rotation within the $xy$ plane, limiting the Chern number to $|C_\pm| = 1$ or $C_\pm = 0$.
For a non-zero Chern number, $d_z(\vec k)$ further needs to change sign as $\vec k$ is varied over the Brillouin zone, i.e., the tight-binding dispersions $\tau_{0}$, $\tau_1$ need to cross,
which is indeed the case for the parameters used in our simulations [Fig.~\ref{fig:berry}(a)].

Alternatively, the Chern number can be computed numerically.
Instead of evaluating Eq.~\eqref{eq:appx-chern} directly, we have performed the computation of the Chern number using the method of Fukui et al. \cite{fukui_chern_2005}.
This indeed yields the Chern numbers of $C_\pm = \mp 1$ for our simulation parameters, in agreement with the geometrical argument given above.
Figure~\ref{fig:berry}(b) shows the the discretized Berry curvature as obtained during the calculation.

\bibliography{consolidated}

\end{document}